\documentclass[
    aps,
    prb,
    longbibliography,
    reprint,       % two-column format
    superscriptaddress, % use superscripts for affiliation
    amsmath,
    amssymb,
    nofootinbib,
    letterpaper,   % set paper size to letter
    showpacs,      % PACS numbers (if required)
    showkeys,      % show keywords (optional)
    floatfix       % better float placement
]{revtex4-2}

\usepackage{graphicx}
\usepackage{subfigure}
\usepackage{booktabs}
\usepackage{amssymb,amsmath}
\usepackage{color}
\usepackage[english]{babel}
\usepackage[utf8]{inputenc}
\usepackage[colorinlistoftodos, color=green!40, prependcaption]{todonotes}
\usepackage{amsthm}
\usepackage{mathtools}
\usepackage{physics}
\usepackage[dvipsnames]{xcolor}
\usepackage{graphicx}
\usepackage{adjustbox}
\usepackage{placeins}
\usepackage{csquotes}
\usepackage{bm}
\usepackage{lipsum}
\usepackage[normalem]{ulem}

\colorlet{linkColour}{magenta}
\colorlet{citeColour}{OliveGreen}
\colorlet{urlColour}{cyan}

\usepackage[colorlinks=true]{hyperref}
\hypersetup{
    unicode=false,          % non-Latin characters
    pdftoolbar=true,        % show Acrobat
    pdfmenubar=true,        % show Acrobat
    pdffitwindow=false,     % window fit to page when opened
    pdfstartview={FitH},    % fits the width of the page to the window
    pdftitle={Noisy composition},    % title
    pdfauthor={Hart Goldman},     % author
    pdfsubject={Subject},   % subject of the document
    pdfcreator={Hart Goldman},   % creator of the document
    pdfproducer={}, % producer of the document
    pdfkeywords={keyword1} {key2} {key3}, % list of keywords
    pdfnewwindow=true,      % links in new window
    colorlinks=true,       % false: boxed links; true: colored links
    linkcolor=linkColour, %red,          % color of internal links (change box color with linkbordercolor)
    citecolor=citeColour,        % color of links to bibliography
    filecolor=linkColour,      % color of file links
    urlcolor=urlColour           % color of external links
}

\newcommand{\bfq}{\mathbf{q}}
\newcommand{\bfr}{\mathbf{r}}

\usepackage{comment}

\begin{document}

\title{The Sound of Electrons Shattering: \\ Current Noise Composition Laws for Electron Fractionalization}

\author{Adarsh S. Patri} 
 \affiliation{Department of Physics and Astronomy \& Stewart Blusson Quantum Matter Institute, University of British Columbia, Vancouver BC, Canada, V6T 1Z4}
\affiliation{Department of  Physics, Massachusetts Institute of Technology, Cambridge, MA 02139}
\author{Josephine J. Yu}
\affiliation{Department of Applied Physics, Stanford University, Stanford, CA 94305, USA }
\author{Yi-Ming Wu}
\affiliation{Department of  Physics, Stanford University, Stanford, CA 94305, USA }
\author{ T. Senthil }
\affiliation{Department of  Physics, Massachusetts Institute of Technology, Cambridge, MA 02139}
\author{ Hart Goldman}
\affiliation{School of Physics and Astronomy, University of Minnesota, Minneapolis, MN 55455, USA }

\date{\today} 

\begin{abstract}
We develop a theory of the non-equilibrium current response for metallic systems near quantum critical points where electronic quasiparticles fractionalize, such as systems near continuous metal-insulator transitions or composite Fermi liquid to Fermi liquid transitions. 
Applying a generalized response theory within a Keldysh path integral framework, we derive a non-perturbative current noise composition law,  wherein the total noise is the sum of the noise of each fractionalized constituent (bosonic holons and fermionic spinons), weighted by their respective resistivities.
We demonstrate that the formally derived composition relations can be interpreted in terms of a simple analogy with  resistors in series.
We leverage this composition rule near certain quantum critical points to show that the shot noise can be suppressed in long nanowires as compared to Fermi liquid expectations due to the collusion of quantum criticality with fractionalization.
\end{abstract}

\maketitle

\section{Introduction}

A wide range of observed phenomena depart dramatically from electronic quasiparticle expectations. 
In the fractional quantum Hall effect, electrons exhibit fractionalization into anyons. On the other hand, strange metals exhibit transport behavior that cannot be captured by coherent quasiparticles. 
Some systems, such as the strange metals in heavy fermion materials \cite{stewart_2001_rev_mod_phys_nfl}, and materials near continuous metal-insulator transitions~\cite{kanoda_PhysRevLett.95.177001,furukawa2018quasi,tingxinli2021, ghitto2021cmit} 
have even invited radical proposals~\cite{lohneysen2007fermi,gegenwart2008quantum,si2001locally,Senthil2004} involving a discontinuous change of the Fermi surface, perhaps associated with electron fractionalization~\cite{Senthil2008,Senthil2004}.
An entirely different class of unusual metals is the (anomalous) composite Fermi liquid~\cite{hlr1993,halperinhalffulllandau2020,dong2023composite, goldman2023composite} that arise in half-filled Landau levels or Chern bands. 
Establishing a clear window into these phenomena calls for new kinds of experimental probes that are capable of resolving the active low-energy degrees of freedom.

\begin{figure}[t!]
\begin{centering}
\includegraphics[width = 0.45\textwidth]{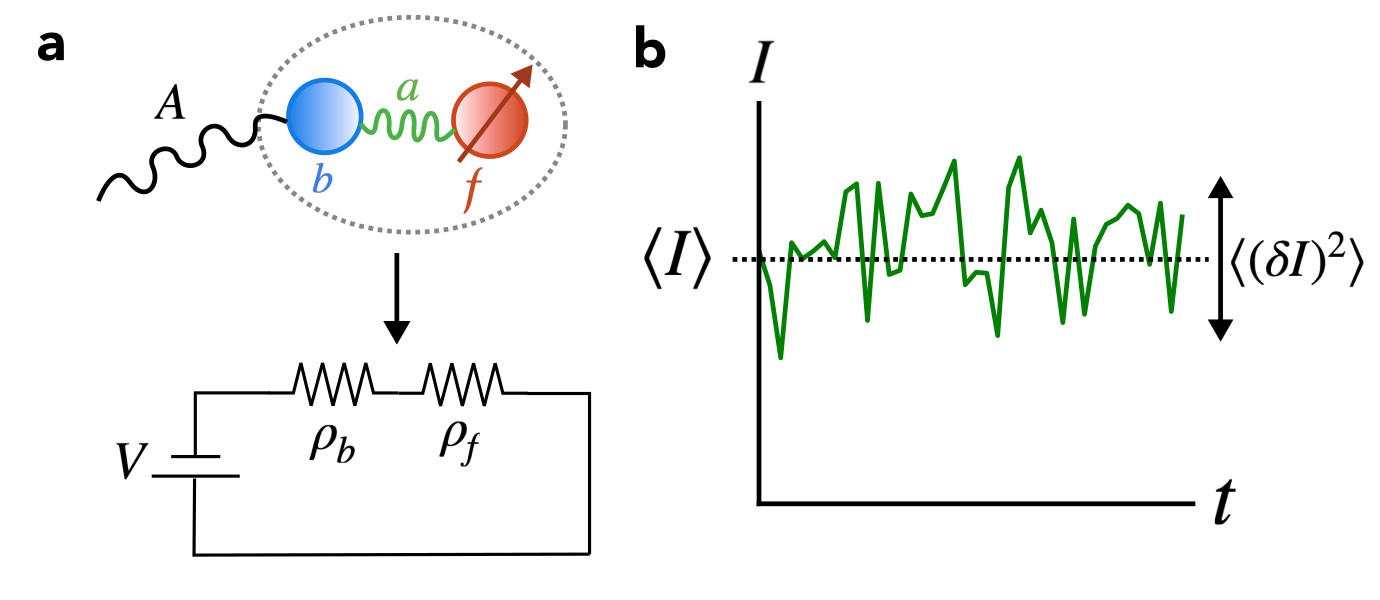}
\end{centering}
\caption{(a) Electronic quasiparticles  coupled to an external EM field $A$ fractionalize into neutral fermionic spinons, $f$, and bosonic charge-carrying holons, $b$,  interacting through an emergent $\mathrm{U}(1)$ gauge field, $a$. 
The key results of this work can be understood via a circuit analogy with resistors in series, with respective resistivities $\rho_f$ and $\rho_b$ corresponding to spinons and holons. 
(b) Schematic plot of measured current $I$ as a function of time $t$, showing the first moment (average current) and second moment (current noise).
\label{fig:schematic}}
\end{figure}

Shot noise is one out-of-equilibrium observable sensitive to the fate of quasiparticles. 
{This is formally defined in terms of the zero-temperature current noise at a finite voltage bias,}
$S \sim \int dt\langle \delta I(t) \delta I(0) \rangle$, and measures fluctuations of the current around its mean (see Fig.~\ref{fig:schematic})~\cite{Datta_1995}, {which in practice can be extrapolated from
$S_{\rm shot}(V,T) = S(V,T)-S(0,T)$}.
It thereby encodes information about the structure of the charged degrees of freedom and their interactions. For example, shot noise has been successfully used to identify the dominant quasiparticle scattering mechanisms in conventional metals \cite{BLANTER20001, kobayashi2001}, to measure the Cooper pair charge in normal metal-superconductor junctions~\cite{jehl2000}, and to identify preformed pairs in the pseudogap phase of cuprate heterostructures \cite{Zhou2019}.
Beyond ordinary electronic quasiparticles, shot noise has also been used to probe the fractional charges in quantum Hall systems~\cite{depicciotto1997,dolev2008, Hashisaka2015}. 

Most recently, the observation of suppressed shot noise in the strange metallic regime of YbRh$_2$Si$_2$~\cite{Chen2023} has been interpreted as indicating quasiparticle demise. 
This observation has motivated much theoretical work seeking to establish expectations for shot noise suppression in metallic systems in which quasiparticle behavior is destroyed by quantum critical fluctuations~\cite{Nikolaenko2023, Wu2024b, Wu2024a, Wang2024,Wang2024a,Wang2025}. However, developing a general understanding of shot noise near metallic quantum critical points (QCPs) remains an open challenge.

In this work, we develop constraints on shot noise for a class of models of metallic critical points associated with the sudden death of an electronic Fermi surface. This model class invokes fractionalization of the electron as a driver of such a continuous phase transition. We note that a sudden reconstruction of the Fermi surface has been postulated to underlie the strange metal physics of YbRh$_2$Si$_2$~\cite{lohneysen2007fermi,gegenwart2008quantum,si2001locally,Senthil2004,RevModPhys.92.011002}.
Specifically, we consider models where electronic quasiparticles, $c$, fractionalize into charged bosonic holons, $b$, and neutral fermionic spinons, $f$, as $c\sim b^\dagger f$, which interact with one another through an emergent $\mathrm{U}(1)$ gauge field (see Fig. \ref{fig:schematic})~\cite{lee_nagaosa_wen_revmod}.
Such decompositions have been especially effective as descriptions of exotic, continuous quantum phase transitions beyond the Landau paradigm out of metallic phases. 
One prominent example is the proposed continuous metal-insulator transition, in which a metallic state ($\langle b\rangle\neq 0$) evolves to an electric insulator ($\langle b\rangle =0$) with a neutral spinon Fermi surface~\cite{Senthil2008}. Similar decompositions have been leveraged for theories of the normal state of cuprates \cite{lee_nagaosa_wen_revmod},
heavy-fermion criticality \cite{Senthil2004, Pepin2007}, and for quantum Hall phase transitions in $2d$ materials\cite{Barkeshli2012,Song2024}.

We develop a theory of the non-equilibrium current response of systems in which electrons decompose into holons and spinons. This approach provides a concrete footing to examine shot noise in a correlated electronic system lacking conventional quasiparticles. Our central result is a non-perturbative composition rule for noise, analogous to the famous Ioffe-Larkin rule~\cite{Ioffe1989} expressing the electronic resistivity as the sum of spinon and holon resistivities. 
We find that within linear-response the physical noise at all temperatures is the sum of the noise of each constituent, weighted by resistivity-dependent factors:
\begin{align}
            S = {e^2} \Bigg(  \left|\frac{\rho_f}{\rho_{f}+\rho_b}\right|^2 S_f + \left|\frac{\rho_b}{\rho_{f}+\rho_b}\right|^2 S_b \Bigg), \label{eq:noise-composition-rule-main}
\end{align}
%%%
where $S_{f}$ ($S_b$) and $\rho_{f}$ ($\rho_b$) are the current noise and resistivity of the spinon (holon). 
This composition rule has an intuitive physical interpretation: It is the result one would obtain for two resistors in series, one corresponding to spinons and the other to holons (see Fig.~\ref{fig:schematic}). 

While the composition of the noise structure follows from fractionalization, the answer can deviate strongly from Fermi liquid expectations near a quantum critical point. In particular, Eq.~\eqref{eq:noise-composition-rule-main} implies that the constituent ($f$ or $b$) which dominates resistivity will also dominate the noise signature. We show that, near a clean QCP where $b$ is gapless --- such as near a continuous metal-insulator transition --- the critical fluctuations of the holons, $b$, will dominate resistivity ($\rho_b\gg\rho_f$) and, in turn, the noise. At the transitions of interest in this work, the critical boson contribution will follow from standard scaling arguments.
There will be a thermal correlation length $\xi_T \sim T^{-\frac{1}{z}}$ where $z$ is the dynamical critical exponent, and a length scale associated with the electric field~\cite{Sondhi1997b} $\ell_E \sim E^{-\frac{1}{z+1}}$, and universal properties will be a scaling function $\Phi$ of their ratio.  
Thus the current noise will behave as 
\begin{equation} 
\label{noise_scaling}
S = k_B T \Phi\left(\frac{\xi_T}{\ell_E} \right) ,
\end{equation}  
where the universal scaling function has the asymptotic behavior $\Phi(x\rightarrow \infty)=x^z$.
In the field-dominated regime  $\ell_E \ll \xi_T$  
the most singular scaling contribution to the noise is
\begin{align}
S\sim S_b &\sim E^{z/(z+1)} +\dots \label{eq:field_dominated_scaling} \\ 
&  \sim E \ell_E +\dots
\end{align}
 We contrast this sub-linear ($z/(z+1)<1$) field-scaling with the noise behavior of a Fermi liquid (FL), $S \sim E L$~\cite{beenakker1992suppression,Nagaev1992a,Nagaev1995}, where $L$ is the length of the wire.
As such, shot noise at criticality is heavily {\it suppressed} as compared to a FL for long wires ($L \gg \ell_E$); on the other hand for wires of length  $L \lesssim  \ell_E$, the critical scaling is {\it enhanced}  compared to its FL counterpart.

The field-dominated regime, $\ell_E \ll \xi_T$, can be recapitulated as,   
\begin{align}
    \frac{V}{T} \frac{\xi_T}{L} \gg 1,
    \label{eq_field_dominated_criterion}
\end{align}
where $V$ is the voltage across the wire.  The critical scaling of Eq.~\eqref{eq:field_dominated_scaling} can be observed in measurements of noise only when Eq.~\eqref{eq_field_dominated_criterion} is satisfied which may be difficult in practice as it requires both a sufficiently high bias and a short nanowire wire. 
However, for long nanowires when Eq.~\eqref{eq_field_dominated_criterion} is not satisfied, the experimentally extracted ``shot noise" $S_{\rm shot}(V,T)$ is actually just a small correction to the equilibrium noise, thus providing an additional route for noise suppression.

\section{Electron decomposition}
The low-energy properties of a variety of correlated electronic systems and their associated continuous phase transitions can be accessed via a parton framework~\cite{wen2004quantum,sachdev2023quantum}, in which the microscopic electron operator is decomposed into constituent particles: a charge-$e$ spin-less boson (`holon', with annihilation operator $b$) and a charge-less spin-1/2 fermion (`spinon', with annihilation operator $f$), $c = b^\dagger f$. Here, $f$ and $b$ are not independent; rather, they are joined by an emergent internal $U(1)$ gauge field $a$, which enforces a matching between the Hilbert spaces of the new representation (in terms of $f$ and $b$) and the original (in terms of $c$). 
 {In other words, the internal gauge field is retained to be dynamical throughout our work, so that the only local (i.e. gauge invariant) operators are the electrons and their composites.}
As depicted in Fig. \ref{fig:schematic}(a), we assign physical charge-$e$ to the $b$, so only the holon couples to the external electromagnetic gauge field ($A$).

A system decomposed in this manner generically includes 
spinon-holon interactions, but these are irrelevant near criticality  in several examples~\cite{Senthil2004,Senthil2008}. For the remainder of this work, we thus consider {a Lagrangian of} the matter sectors minimally coupled to the internal and external gauge fields:
\begin{align}
    \mathcal{L} = \mathcal{L}_b + \mathcal{L}_f = 
    \mathcal{L}[b, A+a] + \mathcal{L}[f, a].
\label{eq_general_parton}
\end{align}
A variety of electronic states are captured by this decomposition, differentiated by the state of the holon $b$ at the mean-field level. For example, holon condensation ($\expval{b}\neq0$) corresponds to a state with mobile charge and a Fermi surface inherited from the spinons $f$; this is the Fermi liquid (FL). Similarly, a gapped holon phase ($\expval{b} =0$) with a residual spinon Fermi surface corresponds to a spin-liquid Mott insulator. Finally, a holon in a bosonic $\nu = 1/2$ Laughlin state (with the spinon in its mean-field FL state) corresponds to the composite Fermi liquid (CFL). The advantage of this decomposition is in its ability to describe not only these phases but also the transitions between them, which can often be expressed in the form Eq.~\eqref{eq_general_parton}. For concreteness, we focus on two examples of such transitions. First, we consider the two-dimensional continuous metal-insulator transition (CMIT). The critical theory of this transition is a superfluid-insulator transition of the charge-sector --- coupled to the spinon Fermi surface --- with a dynamical critical exponent $z = 1$ \cite{Florens2004, sslee2005, Senthil2008}. Second, we consider a composite Fermi liquid to Fermi liquid (CFL-FL) transition, which corresponds to a $\nu = 1/2$ Laughlin-superfluid transition of $b$ with  $z = 1$ \cite{Barkeshli2012,Song2024}.

\section{Response functions from the Keldysh path integral}  
The in- and out- of equilibrium electromagnetic response of a system --- including the current noise --- can be systematically treated within the Keldysh path integral framework.
Within this real-time formalism, path integration is performed over the closed-time loop contour $\mathcal{C}$, where time runs from $t=-\infty$ to $t=\infty$ and back; field variables now exist separately on the forward ($+$) and backwards time branches ($-$), which are manifestly unequal in a non-equilibrium setting~\cite{Rammer2007,Kamenev2023a}.

To extract the current-response functions within this formalism, we consider a generic action of matter fields ($\phi$) minimally coupled to a background electromagnetic field ($A$),
\begin{align}
   & \hspace{-10mm} \mathcal{S} 
    = \int_{\bf{r}}\int_{\mathcal{C}} \mathcal{L}[\phi] - e{j}(x) \cdot A(x) \\
      =  
   \int_{x} &
   \mathcal{L}[\phi^{\rm cl}, \phi^{\rm q}]  - 2e \Big[j^{\rm q}(x) \cdot A^{\rm cl}(x) +j^{\rm cl}(x) \cdot A^{\rm q}(x)  \Big] ,
\end{align}
where $j$ is the current density operator, $e$ is the charge of the matter fields, $x=({\bfr},t)$ is the generalized space-time coordinate, $\int_{x} \equiv \int d^d\bfr\int_{-\infty}^{\infty}dt$. 
In the second line, the fields on the $\pm$ branches are rotated into the ``Keldysh basis" of the classical ($\rm cl$) and quantum ($\rm q$) components. The classical component of the background field $A^{\rm{cl}}$ should be interpreted as the applied driving field, which we will assume is a static, uniform electric field; the quantum component $A^{\rm{q}}$ is the probe or path-integral source field, which generates the current correlation functions. This separation of driving and probe fields allows the Keldysh formalism to capture non-equilibrium response --- a system in a steady state set by $A^{\rm cl}$ can be probed by $A^{\rm q}$. 
Indeed, the mean (average current) and variance (current noise) of the electric current density 
are extracted by taking functional derivatives of the partition function $Z[A^{\rm cl},A^{\rm q}] = \int D[\phi^{cl}, \phi^{\rm {\rm q}}]e^{i\mathcal{S}}$ with respect to the quantum components of the external field,
\begin{align}
  J_n (k) &=   e \langle j^{\rm cl}_n(k) \rangle = \frac{i}{2}\frac{\delta}{\delta A^{\rm q}_n (-k)} \log Z[A^{\rm cl},A^{\rm q}] \bigg |_{A^{\rm q} = 0}, \hspace{-1mm} \label{eq_avg_current} \\
    S_{m n}(k) &= e^2 \langle \delta {j}^{\rm cl}_{m} (k)  \delta {j}^{\rm cl}_n (-k) +  \delta {j}^{\rm cl}_{n} (-k)  \delta {j}^{\rm cl}_{m} (k)    \rangle \nonumber \\
    & = \frac{-1}{2} \frac{\delta}{\delta A^{\rm q} _m(-k)} \frac{\delta}{\delta A^{\rm q}_n (k)}  \log Z[A^{\rm cl},A^{\rm q}] \bigg |_{A^{\rm q} = 0} \hspace{-2mm}.  \label{eq_noise_current}
\end{align}
We employ the compact Fourier notation of $k = (\textbf{k},\omega)$, and $\delta j_\alpha^{\rm cl} (k) = j_\alpha ^{\rm cl}(k) - \langle j_\alpha^{\rm cl} (k) \rangle$ is the deviation of the current density from its average value. The Latin labels $\{m,n\}$ run over the spatial indices. 
Note that only $A^{\rm q}$ is ultimately set to zero, allowing evaluation of the average current $J_n(k)$ and  the current noise $S_{mn}(k)$ both in and out of equilibrium. For example, an explicit evaluation of the noise $S_{mn}$ in equilibrium recovers the standard Johnson-Nyquist noise, while $S_{mn}$ is solely the shot noise in the zero-temperature limit.
Integrating out the dynamical fields $\phi^{\rm{cl}}$ and $\phi^{\rm{q}}$ in the partition function and taking the functional derivatives in Eqs. \eqref{eq_avg_current}, \eqref{eq_noise_current} yields the expressions for the average current and current noise,
\begin{align}
    & J_n(k) = - e^2\Big[\Pi_R(k) \Big]_{nm} A_m^{\rm cl}(k) \label{eq:pi_R_def} \\
    & S_{mn}(k)  = - \frac{ie^2}{2} \Big[ \left(\Pi_K(k)\right)_{mn} + \left(\Pi_K(-k)\right)_{nm}\Big],\label{eq:pi_K_def}
\end{align}
where we have truncated the moments at quadratic order in $A$ (in the expansion of the partition function), absorbed the momentum-dependent constant in the definition of the current density into the Green functions for brevity; the Keldysh Green function is $G_K(q) \equiv -i \expval{\phi^{\rm cl}(q) \bar{\phi}^{\rm cl}(q) }$, and $\Pi_{R/K}$ are the retarded/ Keldysh component of the polarization functions. 
In equilibrium, $\Pi_R$ is related to the conductivity, as in usual linear response theory. However, $\Pi_R$ is well-defined even away from equilibrium --- it can be viewed as the first-order response of a steady-state (set by $A^{\rm cl}$) to a probe field ($A^{\rm q}$). We also see that the current noise $S_{mn}$ is directly related to the Keldysh component of the polarization function, $\Pi_K$, and can therefore be extracted from an effective action expanded to quadratic order in $A$.

\section{Generalized Ioffe-Larkin composition rules for transport}
Equipped with this formalism, we now present the effective response functions for decomposed systems of the form of Eq.~\eqref{eq_general_parton}, keeping in mind our physical examples of the CMIT and CFL-FL transitions. These response functions are extracted from 
the quadratic action for $A$,
\begin{widetext}
\begin{align}
\label{eq_action_effective_A_main_text}
 \hspace{-2mm}   \mathcal{S}_{\rm{eff}}  = -{e^2} & \int_k \Bigg\{A^{\rm q}_{-k} \cdot \left[ \Pi^b_R \cdot \left(\Pi^{f+b}_R \right)^{-1}\cdot \Pi^b_R\right]_k \cdot A^{\rm cl}_k 
    + A^{\rm cl}_{-k} \cdot \left[ \Pi^b_A \cdot \left(\Pi^{f+b}_A \right)^{-1}\cdot \Pi^b_A\right]_k \cdot A^{\rm q}_k - 
\begin{bmatrix}
A^{\rm cl}_{-k} & A^{\rm q}_{-k}
\end{bmatrix}
\begin{bmatrix}
0 & \Pi_A^b \\
\Pi_R^b & \Pi_K^b
\end{bmatrix}_k
\begin{bmatrix}
A^{\rm cl}_k \\
A^{\rm q}_k
\end{bmatrix}\Bigg\} \nonumber\\
     + e^2 &  \int_k A^{\rm q}_{-k} \cdot \left[ \Pi^b_R \cdot \left(\Pi^{f+b}_R\right)^{-1} \cdot \Pi^{f+b}_K \cdot \left(\Pi^{f+b}_A\right)^{-1} \cdot \Pi^b_A - \Pi_K^b \cdot (\Pi_A^{f+b})^{-1} \Pi_A^{b} - \Pi_R^b \cdot (\Pi_R^{f+b})^{-1} \Pi_K^{b}\right]_k \cdot A^{\rm q}_k+\dots, \hspace{-8mm} 
\end{align}
\end{widetext}
which is derived by integrating out the matter fields ($b$ and $f$) and the internal gauge field ($a$) in the partition function.
Here $\Pi^{b/f}$ is the polarization function of the boson/fermion sectors, R/A/K refer to the retarded/advanced/Keldysh components, $\Pi^{f+b} \equiv \Pi^f + \Pi^b$, and we use the notation 
$[f \cdot g \cdot h]_k \equiv [f(k) \cdot g(k) \cdot h(k)]$ and $A^{\alpha}_k \equiv A^{\alpha}(k)$; 
matrix notation is used for the all quantities for compactness -- see Appendix~\ref{app:full-composition-derivation} for derivation details.
 {We note that the irrelevance of the spinon-holon interactions in Eq. \ref{eq_general_parton} is not evoked in this effective action; in fact such interactions are implicitly included in the polarization functions.}

From Eq.~\eqref{eq:pi_R_def}, we have that the average current is determined by the retarded component of the polarization function. Thus, we can immediately identify the physical average current from Eq.~\eqref{eq_action_effective_A_main_text} as
\begin{align}
    {J}_n(k)  
    = -\frac{e^2}{2} & \left[ \Pi^b_R \cdot \left(\Pi^{f+b}_R \right)^{-1}\cdot \Pi^f_R \right]_{k; \ nm} A^{\rm cl}_{m}(k) \\
    - \frac{e^2}{2} & A^{\rm cl}_{m}(k) \left[ \Pi^b_R \cdot \left(\Pi^{f+b}_R \right)^{-1}\cdot \Pi^f_R \right]_{-k; \ mn}^* \nonumber,
\end{align}
where we exploit the relation $\Pi_A(k) = [\Pi_R(k)]^*$.
To provide some intuition for this generalized response function, we consider a system in equilibrium with vanishing Hall components in the polarization function, in which case the average current reduces to
\begin{align}
  \hspace{0mm}  {J}_n({k})
  & = -e^2 \left[  \frac{\Pi^b_R \Pi^f_R}{\Pi_R^{f}+\Pi^{b}_R} 
   \right]_{k;\ nm}A^{\rm cl}_m(k) ,
   \label{eq:conductivity-composition-rule}
\end{align}
where we presume equilibrium polarization functions that obey $[\Pi_R(-k)]^* = \Pi_R(k)$. This produces the established equilibrium Ioffe-Larkin composition rule \cite{Ioffe1989}: 
$\sigma^{-1} = \sigma_f^{-1} + \sigma_b^{-1}$, where $\sigma^{f/b}(k) = \frac{i}{\omega} \Pi_R^{f/b}(k)$, which relates the physical conductivity to the conductivity of each constituent.
Here, conductivity is characterized by the response of the subsystem to the fields it is coupled to. Thus, the resistivity of the fermionic and bosonic subsystems add in series, with the measured resistivity dominated by the more resistive subsystem.

Similarly, the physical current noise of the theory Eq.~\eqref{eq_action_effective_A_main_text} can be identified by comparison with Eq.~\eqref{eq:pi_K_def},
\begin{align}
\hspace{-1mm}    S_{mn} (k) &= -\frac{ie^2}{2} \left[ \Pi^b_R \cdot \left(\Pi^{f+b}_R\right)^{-1}\cdot \Pi^f_K \cdot\left(\Pi^{f+b}_A\right)^{-1} \cdot \Pi^b_A  \right. \nonumber \\ 
    & \hspace{3mm} \left. +   \Pi^f_R \cdot \left(\Pi^{f+b}_R\right)^{-1} \cdot \Pi^b_K \cdot \left(\Pi^{f+b}_A\right)^{-1} \cdot \Pi^f_A  \right]_{k; \ mn}  \nonumber \\
    &  \hspace{5mm} + \Big( m \leftrightarrow n; k \rightarrow - k \Big) \label{eq:full-noise-comp} .
\end{align}
To obtain intuition for the noise composition rule, we focus on the noise along spatial direction $j = \hat{\mathbf{r}}_j$ and again take the case when the Hall components of the polarization functions vanish, leading to a zero-frequency noise power,
\begin{align}
\hspace{-3mm}    S_{jj}(k=0) = {e^2} \Bigg[  \left|\frac{\rho^f}{\rho^{f}+\rho^b}\right|^2 S^f + \left|\frac{\rho^b}{\rho^{f}+\rho^b}\right|^2 S^b \Bigg]_{jj}, 
    \label{eq:noise-composition-rule}
\end{align}
where $S^{f/b} \equiv -i \Pi^{f/b}_K$.
This is the central result of this work: the physical current noise is a sum of the noise of each parton sector, 
weighted by a factor related to the corresponding parton's resistivity. 
We note that these conductivity and noise composition rules can also be derived from a ``kinematic constraint'' on the average motion of the holons $b$ and spinons $f$, as shown in Appendix~\ref{app:composition-via-constraint}.

\section{Circuit interpretation of composition rules} We now offer a simple physical analogy for interpreting our results: both the conductivity, Eq.~\eqref{eq:conductivity-composition-rule}, and noise, Eq.~\eqref{eq:noise-composition-rule}, composition rules can be understood by viewing the system as a circuit with two resistors (corresponding to the boson $b$ and spinon $f$) in series, as sketched in Fig.~\ref{fig:schematic}(a). 
This interpretation is well-established for the conductivity \cite{Ioffe1989}: inverting the conductivity in Eq.~\eqref{eq:conductivity-composition-rule} yields $\rho = \rho_b + \rho_f$, where $\rho=1/\sigma$. The physical resistivity $\rho$ is then the sum of the constituent resistivities $\rho_f$ and $\rho_b$, much like the effective resistance of a circuit with resistors in series is the sum of the individual resistances.

Turning to noise, an analogous composition rule for both voltage and current fluctuations can be extracted from a similar resistors-in-series prescription, wherein each component (resistor) is treated in isolation. For a closed circuit, the total voltage drop, $V$ is a sum of the voltage drop across each component $V=V_1+V_2$; this simple relationship holds for instantaneous voltage fluctuations as well: $\delta V=\delta V_1+\delta V_2$. With a voltage noise defined analogously to current noise, $S_V = \int dt\ \expval{\delta V(t) \delta V(0)}$, the total voltage noise is decomposed into voltage noise across its components in series,
\begin{equation}
    S_V = S_{V,1} + S_{V,2},
    \label{eq:voltage-noise-sum}
\end{equation}
for two independent resistors ($\expval{\delta V_1 \delta V_2} = 0$). 
 Recalling that the voltage noise $S_V$ (at fixed current) and current noise $S_I$ (at fixed voltage) are related by $ S_V = S_I\cdot R^2$ \cite{BLANTER20001} leads to transforming the voltage noise composition rule, Eq.~\eqref{eq:voltage-noise-sum}, into a current noise composition rule,
\begin{equation}
    S = S_1 \left(\frac{R_1}{R} \right)^2 + S_2 \left(\frac{R_2}{R} \right)^2.
    \label{eq:noise-comp-circuit}
\end{equation}
This agrees with the result in Eq.~\eqref{eq:noise-composition-rule} once we take the reasonable assumption that the geometric factors relating $\rho$ to $R$ are identical for both resistors, as the constituent particles experience the same spatial geometry in the physical sample. The circuit interpretation of Eq.~\eqref{eq:noise-composition-rule} is thus that the total current noise follows that of a circuit with two resistors in series: sum of current noise of each resistor \textit{in isolation} ($S_f$ or $S_b$), weighted by a factor related to its resistivity.
These arguments are straightforwardly generalized to the Hall channel of noise ($S_{xy}$), as shown in Appendix~\ref{app:matrix-ressitor-analogy}.
We note that Eq.~\eqref{eq:noise-comp-circuit} relies on the resistors being independent. This apparent independence in Eq.~\eqref{eq_action_effective_A_main_text} (up to quadratic order in $A$) is due to the lack of any direct coupling between the spinon $f$ and boson $b$ in theories of the form Eq.~\eqref{eq_general_parton}.

\section{Consequences}
We now explore the consequences of the noise composition relationship Eq.~\eqref{eq:full-noise-comp} in two examples within the parton framework: the CFL-FL and CMIT transitions. These transitions inherit their critical properties directly from the holon $b$. Thus, $\rho_f$ is expected to be smoothly varying at the critical point, and in the low disorder limit, the spinon sector's contribution to both total noise and total resistivity is expected to be depressed ($\rho_f\ll h/e^2$).  

We now demonstrate how noise near criticality in CMIT and CFL-FL transitions can deviate strongly from FL expectations. 
Recall that the CMIT and CFL-FL transitions share similar critical properties: both transitions are driven by condensation of the bosons, and are characterized by dynamical critical exponent $z=1$. In both cases, the boson resistivity is a finite universal multiple of $h/e^2$~\cite{Senthil2008,Song2024} at the critical point. In the clean limit, the noise is thus solely due to the boson sector,
i.e., $S = e^2 S_b$. The noise is thus given by the scaling form of Eq.~\eqref{noise_scaling} with $z = 1$. 

%%%
In the field-dominated regime $\ell_E \ll \xi_T$, we thus have 
\begin{align}
    S_b  \sim \sqrt{E},
    \label{eq:noise-critical-scaling}
\end{align}
%%%
as has been discussed for superfluid-insulator transitions of bosons at integer lattice filling~\cite{Green2006}.
We direct the reader to Appendix~\ref{app:critical-scaling} for details on the dimensional analysis leading to this scaling.
Through our arguments, the same result is now also seen to hold at the CMIT and CFL-FL transitions, which involve the death of an electronic Fermi surface. 
At the CFL-FL transition, there is also finite Hall conductivity, leading to $S_{xy} \sim \sqrt{E}$ as well. 

 As per our earlier discussion, the shot noise in Eq.~\eqref{eq:noise-critical-scaling} is suppressed relative to the linear-in-$E$ scaling of conventional metallic systems for long nanowire lengths $L \gg \ell_E$. 
We recall that the shot noise can be extracted as {$S_{\rm shot}(V,T) = S(V,T)-S(0,T)$}, where the equilibrium noise is subtracted by the second term. Thus, in the regime where Eq.~\eqref{eq_field_dominated_criterion} is violated, there is a generic suppression of shot noise at these QCPs: the fermionic contribution is also suppressed by $|\rho_f/\rho_b|^2 \ll 1$, and the bosonic contribution consists of small corrections of order $\mathcal{O}((E/T^2)^2)$ to the equilibrium noise.

\section{Discussion}
The derived generalized composition rules for average current and current noise are broadly applicable to a family of strongly-correlated electronic systems that admit a parton description.
Importantly, our results apply to transitions involving the continuous destruction of an electronic Fermi surface; in approaching the critical point from the Fermi liquid, the \textit{electronic} quasiparticle dies continuously but leaves behind a critical Fermi surface of the fermions $f$.  
Focusing on two such models (CMIT and CFL-FL) in 2D, we have demonstrated that, at the resulting QCPs, the current noise in sufficiently long wires is suppressed in comparison to Fermi liquid expectations. Meanwhile, in sufficiently short wires, the zero-temperature current noise about the QCP is instead enhanced and scales sub-linearly with the electric field in the field-dominated regime. 

Furthermore, at both the CMIT and the CFL-FL transition, the scaling of the noise as a function of $T, V, L$ is similar to the bosonic  superfluid-insulator models used to fit the data in YbRh$_2$Si$_2$~\cite{Chen2023}. More generally, we find that the scaling of the noise at these transitions has the same structure as at the superfluid-insulator transition, despite the presence of a critical Fermi surface.
A consequence of this sub-linear scaling is that (in the field-dominated regime) the electric-field dependent Fano factor ($F = S/J$) is also field dependent; when the average current is Ohmic ($J = \sigma E$), $F \sim \sqrt{E}$ due to the lack of scaling of the conductivity in 2D.
Such predictions can be highly relevant to recent observations of continuous metal-insulator transitions in transition-metal dichalcogenide (MoTe$_2$/WSe$_2$) moir\'e superlattices \cite{tingxinli2021}.
Finally, in sufficiently long nanowires, the shot noise can be suppressed due to the total noise being dominated by the equilibrium contributions.

Corrections to the presented boson-sector-dominated critical scaling are expected.
Understanding the nature of these corrections (for instance, by the undamped finite-temperature gauge fluctuations \cite{Witczak-Krempa2012}), the crossover to  Johnson-Nyquist noise, as well as deviations from the clean limit (where the spinons are expected to play an increasingly important role) are important avenues of future study, especially to compare with experiments conducted at finite temperature. Our scaling predictions are valid precisely at criticality, so a careful analysis of the behavior of noise slightly away from the critical point is also an important future direction.

\section*{Acknowledgments}
We are especially grateful to S. Raghu for collaboration at early stages of this work. We also thank S.A. Kivelson for helpful discussions. A.S.P. is supported by NSERC, CIFAR and by the Gordon and Betty Moore Foundation’s EPiQS Initiative through Grant No. GBMF11071 at the University of British Columbia; and was supported by a Simons Investigator Award to T.S. from the Simons Foundation at the Massachusetts Institute of Technology. T.S. was supported by the Department of Energy under grant DE-SC0008739. 
J.J.Y. and Y.-M.W. are supported in part by the US Department of Energy, Office of Basic Energy Sciences, Division of Materials Sciences and Engineering, under Contract No. DE-AC02-76SF00515. J.J.Y. was supported in part by the National Science Foundation Graduate Research Fellowship under Grant No. DGE-1656518. H.G. is supported by startup funds at the University of Minnesota. H.G. is grateful to the Aspen Center for Physics, where some of this work was performed. The Aspen Center for Physics is supported by National Science Foundation grant PHY-2210452.

\bibliography{references}

\onecolumngrid
\appendix
\setcounter{secnumdepth}{1} 
\renewcommand{\thesection}{\Alph{section}}
\renewcommand{\thefigure}{A\arabic{figure}}
\setcounter{section}{0}
\setcounter{figure}{0}

\section*{APPENDIX} 

\section{Derivation of the noise composition rule}
\label{app:full-composition-derivation}

In this section, we derive the noise composition rule Eq.~\eqref{eq:full-noise-comp} in the main text using non-equilibrium quantum field theory (Keldysh contour) methods. The Keldysh framework is well-suited for the examination of response functions of strongly-interacting quantum systems such as those described by Eq.~\ref{eq_general_parton}, unlike the typical Landauer scattering theory (Ref. \cite{Datta_1995}) and kinetic theory \cite{Nagaev1992a,Nagaev1995}.

In a general non-equilibrium system, the time evolution of an observable operator is non-adiabatic.
As such, the standard equilibrium method (either at zero temperature or at finite temperature Matsubara methods) --- where the time evolution on the forward and backward branch is reduced to forward evolution --- is not directly amenable to the problem at hand \cite{Kamenev2023a}.
Instead, the Keldysh technique captures this non-adiabaticity through an explicit integration along the closed-time contour ($\mathcal{C}$) from $t=-\infty$ to $t=\infty$ and back. 
In the continuum notation, the fields along the forwards ($t: -\infty\rightarrow+\infty$) and backwards ($t: +\infty\rightarrow-\infty$) branches are treated as independent measures in the path integral measure, with implicit knowledge of cross-correlations between the forward and backward field components. We direct the reader to Ref. \cite{Kamenev2023a} for a detailed reference on this formalism.

The family of parton theories we consider in this study belong to the real-time action of the form,
\begin{align}
\mathcal{S} &= \mathcal{S}_0 + \mathcal{S}_{a_0} + \mathcal{S}_{a_j}.
\end{align}
Our results do not depend on the specifics of the parton theory, so long as it is of the form Eq.~\ref{eq_general_parton}. However, for clarity, we will use the example of a 2D theory with $z=1$ in the boson sector.
In this case, the action is:
\begin{align}
\mathcal{S}_0&= \int_\mathcal{C} dt\ d^2x \left[ \bar{f}(x,t) G_0^{-1} f(x,t) + \bar{b}(x,t) D_0^{-1} b(x,t) + { V_f(\bar{f}, f) + V_b(\bar{b}, b) }\right] \\ 
\mathcal{S}_{a_0} &= \int_C dt\ d^2x \left[ - ia_0 \bar{f}f  + 2(a_0+eA_0) \omega \bar{b}b  -(a_0+eA_0)^2 \bar{b}b   \right] \\
\mathcal{S}_{a_j} &= \int_\mathcal{C} dt\ d^2x \bigg[ - i a_j(x,t)c_f \left(  (\bar{f}( \partial_j f )-( \partial_j \bar{f}) f) \right) - i (a_j+eA_j) c_b \left(\bar{b}( \partial_j b )-( \partial_j \bar{b}) b)\right) \nonumber\\
 &\quad  - c_f a^2(x,t) \bar{f}(x,t)f(x,t)- c_b (a(x,t)+eA(x,t))^2 \bar{b}(x,t)b(x,t) \bigg] 
\end{align}
where $c_f$ and $c_b$ are constants (involving mass), $V_f$ and $V_b$ are interaction terms for the fermions $f$ and bosons $b$ respectively, and $\mathcal{C}$ represents the closed time contour. We omit the specific structures of the propagators $G_0$ and $D_0$ as they are never invoked in deriving our results.

The $a_0$ component is screened by density-density interactions, and so it mediates only short-range interactions between $f$ and $b$ and does not contribute to the effective low-energy physics.  
The $A_0$ part is screened by the bosons alone, since the spinons do not couple to the external gauge field (are not electrically charged).  The analogous contribution to the propagator of $A_0$ comes from polarization matrix $ \Pi_{00}^b(\omega,q)$; a finite density of bosons will give rise to some screening of $A_0$, so we can neglect this if we are interested in the long-wavelength, low-energy physics. 
We note that there are no cross terms between the temporal $a_0$ and spatial $a_j$ components generated by integrating out the spinons or bosons. This is due to the structure of the polarization imposed by gauge invariance, $q_\mu \Pi_{\mu\nu}^{\alpha\beta} = 0$ (otherwise known as the ``Ward identity") combined with the choice of Coulomb gauge. Indeed, the derivation that follows is valid so long as this decoupling is guaranteed in a given parton theory. In particular, this relationship holds in the $q=0$ and $\omega\rightarrow 0$ limit for the CFL-FL transition as well \cite{Song2024}.
We therefore proceed by neglecting $\mathcal{S}_{a_0}$ because it is screened and gives rise to only short-range interactions.  The low-energy electromagnetic response $\mathcal{S}_{\text{eff}}[A]$ will then not depend on this term. 

Focusing on spatial components of the gauge fields, we rotate all the fields into the Keldysh basis using the same bosonic convention,
\begin{align}
\varphi^1 &= \varphi^{\rm cl} =  \frac{1}{\sqrt{2}}(\varphi^+(t) + \varphi^-(t)) \\
\varphi^2 &= \varphi^{\rm q} = \frac{1}{\sqrt{2}}(\varphi^+(t) - \varphi^-(t)) 
\end{align}
where the bar fields transform in an identical way. 
For ease of notation, we also define the $\gamma$ matrices in the $(cl,q)$ basis:  
\begin{align}
\gamma^{\rm cl} &= \begin{pmatrix}
0 & 1 \\ 
1 &0
\end{pmatrix}, \quad \quad \quad \gamma^{\rm q} = \begin{pmatrix}
1 & 0 \\ 
0 &1
\end{pmatrix}.
\label{eq_keldysh_rotataion_basis}
\end{align}

In this rotated basis, the original action over the Keldysh contour may be written as a single integral on the forward branch ($t=-\infty$ to $t=\infty$):

\begin{align}
\mathcal{S} &=  \int_{-\infty}^\infty dt\ d^dx \bigg[\begin{pmatrix}
\bar{f}^{\rm cl} & \bar{f}^{\rm q}
\end{pmatrix} \begin{pmatrix}
0 & (G_0^{-1})^A \\ 
(G_0^{-1})^R  & (G_0^{-1})^K
\end{pmatrix} \begin{pmatrix}
f^{\rm cl} \\ f^{\rm q}
\end{pmatrix} + \begin{pmatrix}
\bar{b}^{\rm cl} & \bar{b}^{\rm q}
\end{pmatrix} \begin{pmatrix}
0 & (D_0^{-1})^A \\ 
(D_0^{-1})^R  & (D_0^{-1})^K
\end{pmatrix} \begin{pmatrix}
b^{\rm cl} \\ b^{\rm q}
\end{pmatrix}  \nonumber \\ 
&\quad {+ V_f(\bar{f}^{\rm cl},\bar{f}^{\rm q} , f^{\rm cl}, f^{\rm q}) +V_b(\bar{b}^{\rm cl},\bar{b}^{\rm q} , b^{\rm cl}, b^{\rm q}) }\nonumber \\
&\quad   - \frac{ic_f}{\sqrt{2}}  \sum_{\alpha={\rm cl}, {\rm q}} a_j^\alpha \left[ \begin{pmatrix}
\bar{f}^{\rm cl} & \bar{f}^{\rm q}
\end{pmatrix} \gamma^\alpha \begin{pmatrix}
\partial_j f^{\rm cl} \\ \partial_j  f^{\rm q}
\end{pmatrix} - \begin{pmatrix}
\partial_j\bar{f}^{\rm cl} & \partial_j\bar{f}^{\rm q}
\end{pmatrix} \gamma^\alpha \begin{pmatrix}
f^{\rm cl} \\   f^{\rm q}
\end{pmatrix}  \right] \nonumber\\ 
&\quad - \frac{i c_b}{\sqrt{2}}   \sum_{\alpha={\rm cl}, {\rm q}} (a_j+eA_j)^\alpha  \left[ \begin{pmatrix}
\bar{b}^{\rm cl} & \bar{b}^{\rm q}
\end{pmatrix} \gamma^\alpha \begin{pmatrix}
\partial_j b^{\rm cl} \\ \partial_j  b^{\rm q}
\end{pmatrix} - \begin{pmatrix}
\partial_j\bar{b}^{\rm cl} & \partial_j\bar{b}^{\rm q}
\end{pmatrix} \gamma^\alpha \begin{pmatrix}
b^{\rm cl} \\   b^{\rm q}
\end{pmatrix}  \right] \\
 &\quad  - \frac{c_f}{2} \sum_{\alpha\neq \beta}  \begin{pmatrix}
a^{\rm cl} &a^{\rm q}
\end{pmatrix} \gamma^{\alpha}  \begin{pmatrix}
a^{\rm cl}  \\ a^{\rm q}
\end{pmatrix}\begin{pmatrix}
\bar{f}^{\rm cl} & \bar{f}^{\rm q}
\end{pmatrix} \gamma^\beta \begin{pmatrix}
f^{\rm cl} \\ f^{\rm q}
\end{pmatrix} \nonumber \\ 
&\quad - \frac{c_b}{2} \sum_{\alpha\neq \beta}  \begin{pmatrix}
(a+eA)^{\rm cl} & (a+eA)^{\rm q}
\end{pmatrix} \gamma^{\alpha}  \begin{pmatrix}
(a+eA)^{\rm cl}  \\ (a+eA)^{\rm q}
\end{pmatrix}\begin{pmatrix}
\bar{b}^{\rm cl} & \bar{b}^{\rm q}
\end{pmatrix} \gamma^\beta \begin{pmatrix}
b^{\rm cl} \\ b^{\rm q}
\end{pmatrix}\bigg] \nonumber .
\end{align}

The first and second lines contain just the bare actions for the $f$ and $b$ particles. 
We note that the identical choice of the Keldysh rotation (Eq.~\eqref{eq_keldysh_rotataion_basis}) leads to the quadratic part of the boson and fermion actions to have the same structure.
The third and fourth lines describe the coupling between gauge fields and currents, which will give rise to the ``paramagnetic response," which describes current aligning with the applied field.
The last line contributes to the ``diamagnetic response," in which current opposes the applied field.

After Fourier transforming and integrating out the matter fields $f$ and $b$, we expand to quadratic order in the gauge fields $a$ and field $A$  we thus obtain the partition function $Z = \int \mathcal{D}[a] e^{i\mathcal{S}_{\rm eff}[a;A]}$, with the effective action,

\begin{align}
   \mathcal{S}_{\rm eff}[a;A] = \int_q  \left[a_m ^{\alpha}(-q)\right] \Pi_{\alpha \beta; mn}^f (q) \left[a_n^\beta(q)\right] +  \left[a_m ^{\alpha}(-q) + eA_m ^{\alpha}(-q)\right] \Pi_{\alpha \beta; mn}^b (q) \left[a_n^\beta(q) + eA_n^\beta(q) \right] ,
   \label{eq-app:Seff-a-A}
\end{align}
where $\int_q = \int_{-\infty}^\infty d\omega \ d^d \bfq$ and we have defined the polarization matrices for the boson and fermion sectors, $\Pi^{b/f}$. 
We note that, in practice, one cannot integrate out $f$ and $b$ exactly due to the interaction terms; however, the polarization matrices $\Pi^{b/f}$ may be determined within appropriate approximations.
Finally, we integrate out the internal gauge field, $a$, to obtain Eq.~\eqref{eq_action_effective_A_main_text} in the main text.

\section{Kinematic constraints for current and noise}
\label{app:composition-via-constraint}

In this section, we present how the kinematic constraint can be used to derive the conductivity and noise composition rule.
In the parton framework, each physical electron $c$ is decomposed into a charged boson $b$ and a spinon $f$: \begin{equation}
    c^\dagger = b f^\dagger.
\end{equation}
As a consequence, these objects cannot move independently. In particular, each time a spinon hops, a boson must hop as well (in particular, since we define the boson to be a holon, it must hop in the opposite direction). This implies the following constraint on currents
\begin{equation}
J_b  = -J_f.
\label{eq-app:kinematic-constraint}
\end{equation}
Each of these currents may be expressed in the linear response regime as: 
\begin{align}
J_c &= E_c\sigma_c \\ 
J_b &= \epsilon_b \sigma_b \\ 
J_f &= \epsilon_f \sigma_f
\end{align}
where $\epsilon_f = \epsilon$, the ``electric field'' from the internal gauge field, and $\epsilon_b = \epsilon +E$ since the boson is charged both under the external (electromagnetic) field and the internal field. This observation, together with the kinematic constraint, allows us to directly solve for $\sigma_c$ in terms of $\sigma_b$ and $\sigma_f$: 
\begin{align}
 J_b  = -J_f\quad \rightarrow\quad \epsilon = -\frac{\sigma_b}{\sigma_b+\sigma_f} E. 
\end{align}
We also recognize that the physical  current must be carried by the charged objects (in this case, $b$), so 
\begin{equation}
J = J_b \quad \rightarrow\quad \sigma_c = \frac{\sigma_b\sigma_f}{\sigma_b+\sigma_f}.
\end{equation}
We thus recover the well-known result in Eq.~\eqref{eq:conductivity-composition-rule}. 

The above argument gives an idea of how a kinematic constraint relating currents of $f$ and $b$ can be used to derive the composition rules. We now present this derivation, including the origin of this constraint.
For ease of notation we set the gauge charge $e \equiv 1$ in this section.
To derive the composition rules, we relate the response of the system to each of $a$ and $A$; we ``freeze'' $a$ in the path integral and calculate linear response to both the dynamical gauge field $a$ and background gauge field $A$ for a given configuration of the dynamical gauge field $a$. In the spirit of the Keldysh formalism, we also refrain from setting $A^{\rm q}=0$ until the end of the calculation. This proves to be crucial for recovering the noise relation.
The average current with respect to $A$ (the physical current) is: 
\begin{equation}
    -\expval{J_A} = \frac{1}{2}\frac{\delta \mathcal{S}_{\text{eff}}}{\delta A^{\rm q}(-k)}= \Pi^R_b (A^{\rm cl}+a^{\rm cl}) + \Pi^K_b (a^{\rm q}+A^{\rm q})
    \label{eq_J_A_appendix}
\end{equation}

Similarly, we can define a current with respect to $a$. However, since $a$ is dynamical, there are actually two possible currents to consider. The average current with respect to $a^{\rm cl}$ is: 

\begin{equation}
    - \expval{J_a^{\rm cl}} = \frac{1}{2}\frac{\delta \mathcal{S}_{\text{eff}}}{\delta a^{\rm q}(-k)} = (\Pi^R_f+\Pi^R_b)a^{\rm cl} + \Pi^R_b A^{\rm cl} + (\Pi^K_f+\Pi^K_b)a^{\rm q} + \Pi^K_b A^{\rm q},
    \label{eq:app_jc_dynamical}
\end{equation}
and the average current with respect to $a^{\rm q}$ is: 
\begin{equation}
    - \expval{J_a^{\rm q}} = \frac{1}{2}\frac{\delta \mathcal{S}_{\text{eff}}}{\delta a^{\rm cl}(-k)} = (\Pi^R_f+\Pi^R_b) a^{\rm q}  + \Pi^R_b A^{\rm q}.
\end{equation} 
To relate these average currents, we apply the saddle-point approximation for the dynamical gauge field $a$ to analyze the system i.e., $\expval{J_a^{\rm cl}}=\expval{J_a^{\rm q}} = 0$ leading to the equations of motion 
%%%
\begin{align}
    a^{\rm q}&=-\frac{\Pi_b^R}{\Pi_f^R+\Pi_b^R}A^{\rm q} \\
    a^{\rm cl} &= -A^{\rm cl} \frac{\Pi^R_b}{\Pi_b^R + \Pi_f^R} + A^{\rm q}\frac{\Pi_b^R\Pi_f^K - \Pi_b^K \Pi_f^R}{(\Pi_b^R + \Pi_f^R)^2} .
\end{align}
{Note that the saddle-point equations $\expval{J_a^{\rm cl}}=\expval{J_a^{\rm q}} = 0$ reduce to the relation $J_f = -J_b$ when $A^{\rm q}=0$.} Inserting these saddle-point relations into Eq.~\eqref{eq_J_A_appendix}, 
\begin{align}
      -\expval{J_A} = \frac{\Pi_b^R\Pi_f^R}{\Pi_b^R+\Pi_f^R} A^{\rm cl}+ \left[\left(\frac{\Pi_b^R}{\Pi_b^R+\Pi_f^R}\right)^2 \Pi_f^K + \left(\frac{\Pi_f^R}{\Pi_b^R+\Pi_f^R}\right)^2 \Pi_b^K  \right]A^{\rm q},
\end{align}
from which we simultaneously recover the usual composition rule for conductivities Eq.~\eqref{eq:conductivity-composition-rule} (the coefficient of $A^{\rm cl}$) and for the noise Eq.~\ref{eq:noise-composition-rule} (the coefficient of $A^{\rm q}$) in the main text.

\section{Resistor analogy with nonzero Hall response} 
\label{app:matrix-ressitor-analogy}
The resistor analogy for interpreting Eq.~\eqref{eq:noise-composition-rule} may also be used to interpret the full matrix form of the noise composition relation in Eq.~\eqref{eq:full-noise-comp}. 

First, let us rewrite Eq.~\eqref{eq:full-noise-comp} in the limit of $\mathbf{k},\omega \rightarrow 0$ and again exploiting that $\sigma^{f/b}(k) = \frac{i}{\omega} \Pi_R^{f/b}(k)$ and that $\Pi^R_{f/b} = (\Pi_{f/b}^A)^\dagger$: 
\begin{align}
    S^{mn} &= \frac{e^2}{2}\left[ \bm{\rho}_b^{-1} (\bm{\rho}_b^{-1} + \bm{\rho}_f^{-1} )^{-1} \mathbf{S}_f ((\bm{\rho}_b^{-1} + \bm{\rho}_f^{-1} )^{T})^{-1} (\bm{\rho}_b^T)^{-1} + \bm{\rho}_f^{-1} (\bm{\rho}_b^{-1} + \bm{\rho}_f^{-1} )^{-1} \mathbf{S}_b ((\bm{\rho}_b^{-1} + \bm{\rho}_f^{-1} )^{T})^{-1} (\bm{\rho}_f^T)^{-1} \right]^{mn}  \\
    &+ (m\leftrightarrow n) \nonumber
\end{align}
Note that all of the resistivities $\bm{\rho}_f$ and $\bm{\rho}_b$ are real-valued matrices. We insert the identity in the form $\mathbf{1} = \bm{\rho}_f^{-1}\bm{\rho}_f = \bm{\rho}_f^T(\bm{\rho}_f^T)^{-1}$ (and similar for $\rho_b$) and recognize that $(\bm{\rho}_b+\bm{\rho}_f)^{-1} = \bm{\rho}_b^{-1} (\bm{\rho}_b^{-1} + \bm{\rho}_f^{-1} )^{-1}\bm{\rho}_f^{-1}$. Then, we find: 
\begin{equation}
    S^{mn} = \frac{e^2}{2}\left[\bm{\rho}^{-1} \bm{\rho}_f \mathbf{S}_f \bm{\rho}_f^T  (\bm{\rho}^{T})^{-1} + \bm{\rho}^{-1} \bm{\rho}_b \mathbf{S}_b \bm{\rho}_b^T  (\bm{\rho}^{T})^{-1} \right]^{mn} + (m\leftrightarrow n)
\end{equation}
The matrix on the RHS is symmetric under $m\leftrightarrow n$, so we finally have: 
\begin{equation}
    \mathbf{S} = e^2\bm{\rho}^{-1}\left( \bm{\rho}_f \mathbf{S}_f \bm{\rho}_f^T   +  \bm{\rho}_b \mathbf{S}_b \bm{\rho}_b^T   \right) (\bm{\rho}^{T})^{-1}
\end{equation}

We now show that this is exactly the form of the noise matrix from a circuit with resistors (which have nonzero Hall response) in series. From $V_j = I_iR_{ij}$, we have that 
\begin{equation}
    S_V^{nm} = R^{ni} S_I^{ij} (R^{jm})^T.
\end{equation} 

The relationship Eq.~\eqref{eq:voltage-noise-sum} holds component-wise, so 
\begin{equation}
    R^{ni} S_I^{ij} (R^{jm})^T =  R^{ni}_1 S_{I,1}^{ij} (R^{jm}_1)^T+ R_2^{ni} S_{I,2}^{ij} (R^{jm}_2)^T
\end{equation}
where $R^{ij}  = R_1^{ij} + R_2^{ij}$. 

Finally, writing everything in matrix notation, 
\begin{align}
    \mathbf{S} = \mathbf{R}^{-1} \left(\mathbf{R}_1 \mathbf{S}_1 \mathbf{R}_1^{T} + \mathbf{R}_2 \mathbf{S}_2 \mathbf{R}_2^{T}  \right) (\mathbf{R}^T)^{-1}.
\end{align}

The full matrix form of the noise is necessary if one has non-vanishing Hall resistivity of either parton.

\section{Critical scaling for noise and (generalized) conductivity} 
\label{app:critical-scaling}
We present here a description of the dimensional analysis leading to the result Eq.~\eqref{eq:noise-critical-scaling} in the main text.

First, we review the dimensions of current noise $S$ and conductivity $\sigma$ (which has the same dimensions as what we call the ``generalized" conductivity $\tilde{\sigma}$ --- i.e., conductivity beyond standard linear response):
\begin{align}
    S &\sim I^2\cdot t \\ 
    &\sim e^2/t \\ 
    &\sim \epsilon
\end{align}
where $I$ is current and $t$ is time. In the last line, we neglect dimensions of charge, which yields that $S\sim \epsilon$. The analogous analysis for conductivity $\sigma$ is:
\begin{align}
    \sigma &\sim J/E \\ 
    &\sim e/(t \ell^{d-1} E) \\ 
    &\sim e/(t\ell^{d-2} \cdot \ell E) \\
    &\sim \epsilon \cdot e/\ell^{d-2}\cdot 1/\epsilon \\
    &\sim  \ell^{2-d}
\end{align}
where we again neglect $e$. 
Finally, we use scaling near criticality: $1/t \sim 1/\ell^z$ where $z$ is the dynamical exponent~\cite{Sondhi1997b}. Then, since $\epsilon\sim 1/t$, we have $\ell\sim \epsilon^{-1/z}$, this yields: 
\begin{align}
    \sigma &\sim \epsilon^{(d-2)/z}.
\end{align}

Lastly, we go through the argument leading to $\epsilon\sim E^{z/(z+1)}$. Starting from $\epsilon\sim e \ell E$ (the energy gained by a charged particle in an electric field $E$, traversing a distance $\ell$~\cite{Sondhi1997b}), we have: 
\begin{align}
    \epsilon&\sim e \ell E \sim \ell^{-z} 
\end{align}
This allows us to find $\ell$ in terms of $E$ near criticality: 
\begin{align}
    \ell \sim E^{-1/(z+1)}
\end{align}
Finally, we recover:
\begin{align}
    \epsilon&\sim  E^{-1/(z+1)} E \\ 
    &\sim E^{z/(z+1)}.
\end{align}

\end{document}